\documentclass[prd, aps, superscriptaddress, preprintnumbers, twocolumn, floatfix, nofootinbib]{revtex4}

\usepackage{amsfonts}
\usepackage{amsmath}
\usepackage{amssymb}
\usepackage{bm}
\usepackage{dcolumn}
\usepackage{graphicx}
\usepackage[latin1]{inputenc}
\usepackage{latexsym}
\usepackage{rotating}
\usepackage{graphicx}
\usepackage{color}
\usepackage{float}
\usepackage{epsfig}

\usepackage[unicode=true,pdfusetitle,
 bookmarks=true,bookmarksnumbered=false,bookmarksopen=false,
 breaklinks=false,pdfborder={0 0 1},backref=false,colorlinks=true]
 {hyperref}
\hypersetup{
 linkcolor=blue, citecolor=magenta, urlcolor=red, filecolor=blue}

\begin{document}

\title {Point Particle Motion in DFT and a Singularity-Free Cosmological Solution}

\author{Robert Brandenberger}
\email{rhb@physics.mcgill.ca}
\affiliation{Physics Department, McGill University, Montreal, QC, H3A 2T8, Canada}

\author{Renato Costa}
\email{Renato.Santos@uct.ac.za}
\affiliation{Cosmology and Gravity Group, Dept. of Mathematics and Applied Mathematics,
University of Cape Town, Rondebosch 7700, South Africa}

\author{Guilherme Franzmann}
\email{guilherme.franzmann@mail.mcgill.ca}
\affiliation{Physics Department, McGill University, Montreal, QC, H3A 2T8, Canada}

\author{Amanda Weltman}
\email{amanda.weltman@uct.ac.za}
\affiliation{Cosmology and Gravity Group, Dept. of Mathematics and Applied Mathematics,
University of Cape Town, Rondebosch 7700, South Africa}

\date{\today}

\begin{abstract}

We generalize the action for point particle motion to a double field theory background.
After deriving the general equations of motion for these particle geodesics, we
specialize to the case of a cosmological background with vanishing antisymmetric
tensor field. We then show that the geodesics can be extended
to infinity in both time directions once we define the appropriate physical clock. Following this prescription, we also show the existence of a singularity-free cosmological solution.

\end{abstract}

\maketitle

\section{Introduction}

If the fundamental building blocks of matter are elementary superstrings instead of point particles,
the evolution of the very early universe will likely be very different than in Standard Big Bang cosmology.
``String Gas Cosmology'' is a scenario for the very early stringy universe which was proposed some
time ago \cite{BV} (see e.g. \cite{SGCrevs} for some more recent reviews). String Gas Cosmology is
based on making use of the key new degrees of freedom and symmetries which distinguish string
theories from point particle theories. The existence of string oscillatory modes leads to a maximal
temperature for a gas of strings in thermal equilibrium, the ``Hagedorn temperature'' $T_H$ \cite{Hagedorn}.
Assuming that all spatial dimensions are toroidal with radius $R$, the presence of string winding modes leads to
a duality,
\begin{equation} \label{Tdual}
R  \, \rightarrow \, \frac{1}{R},
\end{equation}
(in string units) in the spectrum of string states. This comes about since the energy
of winding modes is quantized in units of $R$, whereas the energy of momentum modes
is quantized in units of $1/R$. The symmetry (\ref{Tdual}) is realized by interchanging
momentum and winding quantum numbers\footnote{See also \cite{Boehm} for an
extended discussion of T-duality when branes are added.}.

As was argued in \cite{BV}, in String Gas Cosmology the temperature singularity of the Big Bang
is automatically resolved. If we imagine the radius $R(t)$ decreasing from some initially
very large value (large compared to the string length), and matter is taken to be a gas of
superstrings, then the temperature $T$ will initially increase, since for large values of $R$ most
of the energy of the system is in the light modes, which are the momentum modes, and
the energy of these modes increases as $R$ decreases. Before $T$ reaches the maximal
temperature $T_H$, the increase in $T$ levels off since the energy can now go into producing
oscillatory modes. For $R < 1$ (in string units) the energy will flow into the winding modes which
are now the light modes. Hence,
\begin{equation}
T(R) \, = \, T\left(\frac{1}{R}\right) \, .
\end{equation}
A sketch of the temperature evolution as a function of $R$ is shown in Figure 1.
As a function of $\ln{R}$ the curve is symmetric as a reflection of the symmetry
(\ref{Tdual}). The region of $R$ when the temperature is close to $T_H$ and the
curve in Fig. 1 is approximately horizontal is called the ``Hagedorn phase''. Its
extent is determined by the total entropy of the system \cite{BV}.

\begin{figure}[h]
    \centering
    \includegraphics[scale = 0.4] {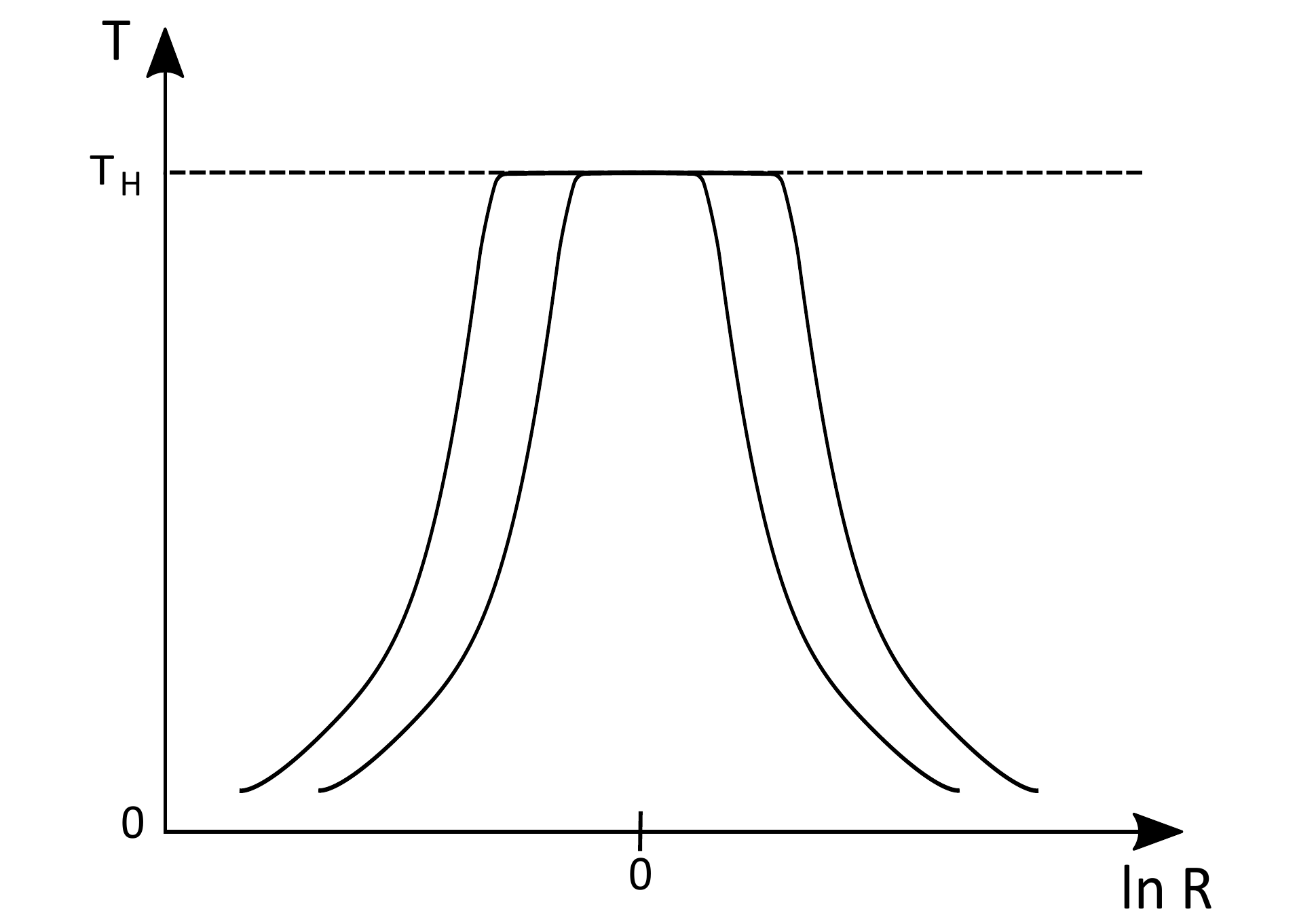}
    \caption{T versus $\log{R}$ for type II superstrings. Different curves are obtained for different entropy values, which is fixed. The larger the entropy the larger the plateau, given by the Hagedorn temperature. For $R=1$ we have the self-dual point.}
    \label{fig:my_label}
\end{figure}

In \cite{BV} it was furthermore argued that at the quantum level there must be
two position operators for every topological direction, one operator $X$ dual to
the momentum number (this is the usual position operator for point particle theories)
and a dual operator ${\tilde X}$ which is dual to the winding number.
The physically measured length $l(R)$ will always be determined by the light
modes of the system. Hence, for large $R$ it is determined by $X$, but for small
$R$ it is determined by ${\tilde X}$. Thus,
\begin{eqnarray} \label{plength}
l(R) \, &=& \, R  \,\,\,\, {\rm for} \,\,\,\, R \gg 1 \, ,\\
l(R) \, &=& \, \frac{1}{R} \,\,\,\, {\rm for} \,\,\,\, R \ll 1 \,  .  \nonumber
\end{eqnarray}

More recently, a study of cosmological fluctuations in String Gas Cosmology
\cite{NBV} showed that thermal fluctuations in the Hagedorn phase of an
expanding stringy universe will evolve into a scale-invariant spectrum of
cosmological perturbations on large scales today (see \cite{BNPV1} for
a review). If the string scale is comparable to the scale of particle physics
Grand Unification the predicted amplitude of the fluctuations matches the
observations well (see \cite{Ade:2015lrj} for recent observational results). The
scenario also predicts a slight red tilt to the scalar power spectrum. Hence, String Gas
Cosmology provides an alternative to cosmological inflation as a theory for
the origin of structure in the Universe. String Gas Cosmology predicts \cite{BNPV2}
a slight blue tilt in the spectrum of gravitational waves, a prediction by means
of which the scenario can be distinguished from standard inflation (meaning
inflation in Einstein gravity driven by a matter field obeying the usual energy
conditions). A simple modelling of the transition between the
Hagedorn phase and the radiation phase leads to a running of the spectrum
which is parametrically larger than what is obtained in simple inflationary
models \cite{Liang}.

What is missing to date in String Gas Cosmology is an action for the dynamics
of space-time during the Hagedorn phase. Einstein gravity is clearly inapplicable
since is does not have the key duality symmetry (\ref{Tdual}). In ``Pre-Big-Bang Cosmology''
\cite{PBB} it was suggested to use dilaton gravity as a dynamical principle since
the T-duality symmetry yields a scale factor duality symmetry. However, dilaton
gravity does not take into account enough of the stringy nature of the Hagedorn
phase.

``Double Field Theory'' \cite{Siegel:1993xq,Siegel:1993th,Siegel:1993bj,DFT} (see e.g. \cite{DFTrev} for a review)
has recently been introduced as a field theory which is
consistent with the T-duality symmetry of string theory. Given a
topological space, in Double Field Theory
(DFT) there are two position variables associated with every direction
of the topological space which is compact. Since DFT is based on the
same stringy symmetries as String Gas Cosmology, it is reasonable
to expect DFT to yield a reasonable prescription for the dynamics of
String Gas Cosmology.

On the other hand, even DFT will not yield an
ideal dynamical principle for String Gas Cosmology since DFT only
contains the massless modes of the bosonic string theory: the metric, the dilaton
and the antisymmetric tensor field. We can try to include the other
stringy degrees of freedom through a matter action in the same way
as done in \cite{BV}. Hence, ultimately we would like to study the cosmological
equations of motion of DFT in the presence of string gas matter.
As a first step towards this goal we will in this paper study point
particle motion in DFT. Through this study we can explore time-like
and light-like geodesics. We will argue that (taking into account the
appropriate definition of time) these geodesics are complete.
This yields further evidence that string theory can lead to a nonsingular
early universe cosmology. In particular, we also show that vacuum DFT background equations of motion produce a singularity-free cosmological solution when this new definition of time is considered.

\section{Essentials of Double Field Theory}

DFT is a field theory which lives in a ``doubled'' space in which the
number of all dimensions with stable string windings is doubled. From
the point of view of string theory this means having one spatial dimension
dual to the momentum, and another one dual to the windings. We
will here consider a setup in which all spatial dimensions have windings.
Thus, our DFT will live in $(2D -1)$-dimensions, where $(D-1)$ is the number
of spatial dimensions of the underlying manifold. Note that under toroidal compactifications, the corresponding T-dualily group  is $O(n,n)$, where $n$ corresponds to the number of spatial compact dimensions. In DFT, the theory is covariantly formulated in the double space, so that the underlying symmetry group is $O(n,n)$ \cite{DFTrev}. Thus, any scalar object should be invariant under this group transformations. This will be relevant for when we build the point particle action in DFT below. We will denote the
usual spatial coordinates by $x^i$ and the dual coordinates by ${\tilde{x}_i}$.

We consider a cosmological space-time in standard General Relativity given by:
\begin{equation} \label{metricGR}
ds^2 \, = \, - dt^2 + g_{ij} dx^i dx^j \, ,
\end{equation}
where $t$ is physical time and $g_{ij}$ is the $(D-1)$-dimensional spatial metric.
The coordinates $i$ and $j$ run over these original spatial indices. In DFT
the metric in doubled space-time (all spatial dimensions doubled, but
not time) is written in terms of a generalized metric $\mathcal{H}_{MN}$,
where $M$ and $N$ run over all $(2D - 1)$ space-time indices:
\begin{eqnarray}
\label{metricDFT}
dS^2 \, = \, \mathcal{H}_{MN}dX^M dX^N \, .
\end{eqnarray}
The generalized metric depends both on the original metric and on
the antisymmetric tensor field $b_{ij}$. In the case of a cosmological
background we will usually separate out the time component and
write the line element as:
\begin{eqnarray}
\label{metricDFTcosmo}
dS^2 \, =  \, - dt^2 + \mathcal{H}_{MN}dX^M dX^N \, ,
\end{eqnarray}
where now $M$ and $N$ run only over spatial indices.
In DFT all massless string states are considered. Hence, in addition to
the metric there is a dilaton $\phi$ and an antisymmetric tensor
field $b_{ij}$.  The generalized metric is then given by:
\begin{eqnarray}
\mathcal{H}_{MN} \, = \,
\begin{bmatrix}
g^{ij} & -g^{ik}b_{kj} \\
b_{ik}g^{kj} & g_{ij}-b_{ik}g^{kl}b_{lj} \\
\end{bmatrix} \, ,
\end{eqnarray}
where the indices are raised with the usual Riemannian metric.

The DFT action is chosen to treat $g_{ij}$ and
$b_{ij}$ in a unified way, and to reduce to the supergravity action
if there is no dependence on the dual coordinates. It is
given by:
\begin{eqnarray}\label{DFT_action}
S \, = \,  \int dxd\tilde{x} e^{-2d}\mathcal{R},
\end{eqnarray}
where $d$ contains both the dilaton $\phi$ and the determinant of the metric,
\begin{equation}
e^{-2d} \, = \, \sqrt{-g} e^{-2 \phi} \, ,
\end{equation}
and where \cite{Hohm:2010jy},
\begin{eqnarray}
\mathcal{R} & = & \frac{1}{8}\mathcal{H}^{MN}\partial_{M}\mathcal{H}^{KL}\partial_{N}\mathcal{H}_{KL}-\frac{1}{2}\mathcal{H}^{MN}\partial_{M}\mathcal{H}^{KL}\partial_{K}\mathcal{H}_{NL}\nonumber \\
 &+&4\mathcal{H}^{MN}\partial_{M}\partial_{N}d-\partial_{M}\partial_{N}\mathcal{H}^{MN}-4\mathcal{H}^{MN}\partial_{M}d\partial_{N}d\nonumber \\
 &+&4\partial_{M}\mathcal{H}^{MN}\partial_{N}d,
\end{eqnarray}
with the matrix $\eta^{MN}$ being given by:
\begin{eqnarray}
\eta^{MN} \, = \,
\begin{bmatrix}
0 & \delta^{\;\;j}_i \\
\delta^{i}_{\;\;j} & 0\\
\end{bmatrix} \, ,
\end{eqnarray}
and we are writing the doubled space coordinates as:
 \begin{equation}
X^M \, = \,(\tilde{x}_i,x^{i}).
\end{equation}
Finally, note that since all the fields now depend on double coordinates, in principle we have doubled the number of degrees of freedom we had started with. In order to eliminate these extra degrees of freedom, one usually considers the section condition \cite{DFTrev}, which eliminates the dual-coordinate dependence of all the fields.

\section{Point Particle Motion in Double Field Theory}

The action for the massive relativistic point particle
with world line coordinates $x^i(t)$ is given by:
\begin{eqnarray}
S \, = \,-mc \int ds \, \label{eq:RelAction} ,
\end{eqnarray}
where the line element $ds$ is given by (\ref{metricGR}).
The natural generalization of it which corresponds to the action of a point particle with world line in doubled space given by $X^M(t)$ in a DFT background
is written in the following way:
\begin{equation} \label{action}
    S = -mc \int dS ,
\end{equation}
where the generalized line element $dS$ is given by (\ref{metricDFT}). This action has also been introduced in \cite{Ko:2016dxa}, although the geodesic equations have not been worked out. Note this action is covariant under $O(n,n)$ transformations, as expected. Moreover, if the section condition is imposed, it recovers (\ref{eq:RelAction}).

Before deriving the geodesic equations, a few comments are in order. The generalized metric is a constrained object, which satisfies:
\begin{equation}
\mathcal{H}\eta\mathcal{H} \, = \, \eta^{-1},
\end{equation}
therefore its variation is constrained as well, as showed in \cite{zwiehohmhull1}, and it is given by:
\begin{align}
\frac{\partial\mathcal{H}_{MN}^{(c)}}{\partial X^{P}}=\frac{1}{4}\left[\left(\eta_{MQ}+\mathcal{H}_{MQ}\right)\frac{\partial\mathcal{H}^{QR}}{\partial X^{P}}\left(\eta_{RN}-\mathcal{H}_{RN}\right)+\right. \nonumber\\
\left.+\left(\eta_{MQ}-\mathcal{H}_{MQ}\right)\frac{\partial\mathcal{H}^{QR}}{\partial X^{P}}\left(\eta_{RN}+\mathcal{H}_{RN}\right)\right],
\end{align}
where the index $(c)$ specifies when we consider constrained objects.

Varying the action with respect to the world sheet coordinates
$X^M(t)$ yields the following equations of motion:

\begin{widetext}

\begin{eqnarray}\label{relativisticDFT}
\mathcal{H}_{MN}\frac{d^2 X^N}{dS^2}+ \frac{\partial \mathcal{H}^{(c)}_{MN}}{\partial X^P}\frac{d X^P}{dS} \frac{dX^N}{dS}-\frac{1}{2} \frac{\partial\mathcal{H}^{(c)}_{PN}}{\partial X^M}  \frac{dX^P}{dS}\frac{dX^N}{dS}=0.
\end{eqnarray}
Then, the equation of motion for the dual coordinates ($M=1$) is:
\begin{eqnarray} \label{eq1}
g^{ij}\frac{d^2 \tilde{x}_j}{dS^2} - g^{ik}b_{kj}\frac{d^2 x^j}{dS^2}
 + (\tilde{\partial}^m g^{in})\frac{d \tilde{x}_n}{dS} \frac{d\tilde{x}_m}{dS} + (\partial_m g^{in})\frac{d x^m}{dS} \frac{d\tilde{x}_n}{dS}+ \tilde{\partial}^m (g^{ik}b_{kn})\frac{d \tilde{x}_m}{dS} \frac{dx^n}{dS}+ \partial_m (g^{ik}b_{kn})\frac{d x^m}{dS} \frac{dx^n}{dS}
 \nonumber \\-\frac{1}{2}\bigg[ (\tilde{\partial}^i g^{mn}) \frac{d\tilde{x}_m}{dS}\frac{d\tilde{x}_n}{dS}-\tilde{\partial}^i (g^{mk}b_{kn})  \frac{d\tilde{x}_m}{dS}\frac{dx^n}{dS}+ \tilde{\partial}^i (b_{mk}g^{kn}) \frac{dx^m}{dS}\frac{d\tilde{x}_n}{dS}+ \tilde{\partial}^i(g_{mn}-b_{mk}g^{kj}b_{jn}) \frac{dx^m}{dS}\frac{dx^n}{dS}\bigg] \, = \, 0 \, ,
\end{eqnarray}
whereas the equation of motion for the regular coordinates ($M=2$) is:
\begin{eqnarray} \label{eq2}
\left(g_{ij}-b_{ik}g^{kl}b_{lj}\right)\frac{d^2x^j}{dS^2}
+\partial_l (g_{in}-b_{ik}g^{kj}b_{jn})\frac{dx^n}{dS}\frac{dx^l}{dS}
-\frac{1}{2} \partial_i(g_{mn}-b_{mk}g^{kj}b_{jn})\frac{dx^m}{dS}\frac{dx^n}{dS}
\nonumber\\
+b_{ik}g^{kj}\frac{d^2 \tilde{x}_j}{dS^2}+\tilde{\partial}^l (b_{ik} g^{kn})\frac{d\tilde{x}_l}{dS}\frac{d\tilde{x}_n}{dS} + \partial_j (b_{ik}g^{kn})\frac{d\tilde{x}_n}{dS}\frac{dx^j}{dS}
+\tilde{\partial}^l (g_{in} - b_{ik}g^{kj}b_{jn})\frac{d\tilde{x}_l}{dS}\frac{dx^n}{dS} \nonumber \\
-\frac{1}{2} \bigg[ \partial_i g^{mn}\frac{d\tilde{x}_m}{dS}\frac{d\tilde{x}_n}{dS} -
\partial_i (g^{mk}b_{kn})\frac{d\tilde{x}_m}{dS}\frac{dx^n}{dS} +
\partial_i(b_{mk}g^{kn})\frac{d\tilde{x}_n}{dS}\frac{dx^m}{dS} \bigg] \, = \, 0.
\end{eqnarray}
These are the most general equations for a point particle in a DFT background with a metric and
a 2-form field. From the first line of equation (\ref{eq2}) it is easy to see that after imposing the section condition and setting the two-form to be zero, we are left with the geodesic equation of a relativistic point particle.

\end{widetext}

\section{Point Particle Motion in a Cosmological Background}

Now we want to specialize the discussion to a homogeneous and isotropic cosmological background
with vanishing $b_{ij}$. We thus consider the cosmological metric:
\begin{equation} \label{metricCOSMO}
ds^2 \, = \, - dt^2 + a^2(t) \delta_{ij} dx^i dx^j + a^{-2}(t)\delta^{ij} d\tilde{x}_i d\tilde{x}_j\, ,
\end{equation}
where $a(t)$ is the scale factor. Setting the antisymmetric tensor field to zero, the general equations of motion of the previous section simplify to:
\begin{eqnarray}
\frac{d}{dS}\left(\frac{d\tilde{x}_{a}}{dS}\frac{1}{a^2}\right) & = & 0\\
\frac{d}{dS}\left(\frac{dx^{a}}{dS}a^{2}\right) & = & 0.
\end{eqnarray}
These are the geodesic equations of point particle motion of DFT in a cosmological
background.

We will now argue that geodesics are complete in the sense that they
can be extended to arbitrarily large times both in the future and in the past. This is
true for all particle geodesics except for the set of measure zero where either
all coordinates $x^i$ or all coordinates ${\tilde x}_i$ vanish. We will consider
a given monotonically increasing scale factor $a(t)$, like the scale factor of
Standard Big Bang cosmology. Note that in this parametrization, the coordinate
$t$ lies in the interval between $t = 0$ and $t = \infty$.

Consider a trajectory at some initial time $t_0$ with the property that some $x^i$ and
some ${\tilde x}_j$ are non-vanishing. Due to Hubble friction, then the velocity
$dx^i / dt$ will decrease. On the other hand, the dual velocity $d{\tilde x}_j / dt$
will approach the speed of light. Hence, the proper distance $\Delta S$ in double space
between time $t_0$ and some later time $t_2$ will be:
\begin{equation}
\Delta S \, = \int_{t_0}^{t_2} \gamma(t)^{-1} (1 + T_2)^{1/2} dt  \, ,
\end{equation}
where $T_2$ is the contribution from the dual which ceases to increase at
late times since the dual velocity goes at the speed of light, and $\gamma(t)$ is
the relativistic $\gamma$ factor of the motion in the $x^i$ direction (for simplicity
we consider motion only in one original direction and in one dual direction).
Hence, the geodesic can be extended to infinite time in the future.

Now consider evolving this geodesic backwards in time from $t_0$ to some
earlier time $t_1$. Then it is the motion of the dual coordinates which comes
to rest. The proper distance in double space is now:
\begin{equation}
\Delta S \, = \, \int_{t_1}^{t_0} {\tilde{\gamma}}(t) (1 + T_1)^{1/2} dt  \, ,
\end{equation}
where ${\tilde{\gamma}}$ is the relativistic gamma factor for motion in
the dual space directions, and $T_1$ is the contribution to the proper distance
which comes from the regular spatial dimensions which is negligible at
very early times since the velocity in the regular directions approaches the
speed of light.

The expansion of the scale factor in the dual spatial directions as time
decreases is analogous to the expansion in the regular directions as
time increases. In line with T-duality we propose to view the dynamics
of the dual spatial dimensions as $t$ decreases as expansion when
the {\it dual time},
\begin{equation}
t_d \, = \, \frac{1}{t},
\end{equation}
increases. In fact, in analogy to the definition of {\it physical length}
$l(R)$ in (\ref{plength}) \cite{BV}, we can define a {\it physical time}
$t_p(t)$ as:
\begin{eqnarray} \label{ptime}
t_p(t) \, &=& \, t  \,\,\,\, {\rm for} \,\,\,\, t \gg 1 \, ,\\
t_p(t) \, &=& \, \frac{1}{t} \,\,\,\, {\rm for} \,\,\,\, t \ll 1 \,  .  \nonumber
\end{eqnarray}
With this definition, the geodesics studied in this paper are
geodesically complete in the sense that they can be extended
in both directions to infinite time.

We can also justify the above argument by dualizing both space
and time, \textit{i.e}., by also introducing a dual time ${\tilde t}$ which is
dual to ``temporal winding modes" of the string \cite{Hull2}. This concept
can be made rigorous in Euclidean space-time where time is
taken to be compact. When the regular Euclidean time domain
shrinks in size, the dual time domain increases. This is analogous
to the dual space domain increasing as $1/R$ when the regular
space domain $R$ is decreasing.  From this point of view, the
definition (\ref{ptime}) is simply the time component of (\ref{plength}).
We can also write (\ref{ptime}) as:
\begin{eqnarray} \label{ptime2}
t_p(t) \, &=& \, t  \,\,\,\, \,\,\,\,\, \,\,\,\, \,\,\, {\rm for} \,\,\,\, t \gg 1 \, ,\\
t_p(t) \, &=& \, {\tilde t} = t_d \,\,\,\, {\rm for} \,\,\,\, t \ll 1 \,  .  \nonumber
\end{eqnarray}

At finite temperatures $T$, the string partition function $Z(T)$ is periodic in
Euclidean time $\beta = 1/T$, and - at least for certain string theory setups -
satisfies the temperature duality:
\begin{equation}
Z(\beta) \, = \, Z\left(\frac{1}{\beta}\right),
\end{equation}
which is a consequence of the T-duality symmetry. Based on this
symmetry it was argued \cite{Costas} that these string theory models
correspond to bouncing cosmologies in which the physical
temperature is taken to be (always in string units):
\begin{eqnarray} \label{pTemp}
T_p(T) \, &=& \, T  \,\,\,\,  {\rm for} \,\,\,\, T \ll 1 \, ,\\
T_p(T) \, &=& \, \frac{1}{T} \,\,\, {\rm for} \,\,\,\, T \gg 1 \,.   \nonumber
\end{eqnarray}
Note that two time formalism based on ideas from string theory were
also discussed in \cite{Hull} and \cite{Bars}.

\section{Singularity-free cosmological background}

In this section, we study a geometrical approach to formalize the idea of the physical clock introduced in the last section and conclude that upon considering this definition of time, the vacuum solutions of the DFT background equations are singularity-free. The DFT cosmological background equations of motion in the presence of a hydrodynamical fluid will be discussed in \cite{DFTbackground}.

We start off considering the following ansatz:
\begin{eqnarray}
dS^2 \, = \, &-&dt^2 -d\tilde{t}^2 + a^2(t,\tilde{t})\sum_{i=1}^{D-1}  dx^i dx^i
\nonumber\\
&+& a^{-2}(t,\tilde{t})\sum^{D-1}_{j=1}  d\tilde{x}^j d\tilde{x}^j,
\end{eqnarray}
in the DFT equations of motion, resulting in \cite{Wu:2013sha}:
\begin{flalign}
&\left[4\partial_{\tilde{t}} \partial_{\tilde{t}}d - 4(\partial_{\tilde{t}}d)^2 - (D-1)\tilde{H}^2\right] +\nonumber \\ &\left[4\partial_t\partial_t d - 4(\partial_td)^2 - (D-1)H^2\right] =0 \\
&\left[-(D-1)H^2 + 2 \partial_t \partial_t d\right] - \left[- (D-1)\tilde{H}^2 + 2 \partial_{\tilde{t}} \partial_{\tilde{t}}d\right] =0 \\
&\left[\dot{\tilde{H}} - 2\tilde{H}\partial_{\tilde{t}} d\right] + \left[\dot{H} - 2H\partial_t d\right] =0,
\end{flalign}
where $H= a^{-1}da/dt$ and $\tilde{H} = a^{-1}da/d\tilde{t}$.

The solution to these equations are given by
\begin{eqnarray}
a_\pm (\tilde{t},t) \, =
\, \left|\frac{t}{\tilde{t}}\right|^{\pm 1/\sqrt{D-1}}, \quad d(t,\tilde{t}) = -\frac{1}{2} \ln{|t\tilde{t}|} \\
a_\pm (\tilde{t},t) \, =
\, \left|t\tilde{t}\right|^{\pm 1/\sqrt{D-1}}, \quad d(t,\tilde{t}) = -\frac{1}{2} \ln{|t\tilde{t}|}.
\end{eqnarray}
However, so far there was no clear interpretation of these equations and solutions given the presence
of the extra time coordinate, $\tilde{t}$.

Within the prescription we have introduced in the last section, we can interpret this extra time coordinate
as the geometrical clock associated to the winding modes\footnote{We will discuss further about this prescription in \cite{DFTbackground}.}. In particular, we have provided
arguments that this clock, when seen from the momentum perspective, should correspond to:
\begin{equation}
    \tilde{t} \, = \, \frac{1}{t}.
\end{equation}
Thus, the above ansatz can be seen as a way to implement the ideas we have introduced at a
geometrical level. By doing so, the effective line element becomes:
\begin{eqnarray}
dS^2 \, = \, &-&\left(1+\frac{1}{t^4}\right) dt^2  +a^2(t) \sum_{i=1}^{D-1}  dx^i dx^i
\nonumber\\
&+& a^{-2}(t)\sum^{D-1}_{j=1}  d\tilde{x}^j d\tilde{x}^j.
\end{eqnarray}
The solutions of the DFT equations of motion become:
\begin{align}
&a_\pm (t) \, = \, |t|^{\pm 2\sqrt{D-1}}, \quad &d(t) = \text{const.} \\
&a_\pm (t) \, = \, \text{const.}, \quad &d(t) =\text{const.} \label{eq:non_trivial}
\end{align}

Now, using instead the physical clock, we can rewrite the line element by considering the following
\begin{equation}
dt_{p} \, = \, \sqrt{1+\frac{1}{t^{4}}}dt,
\end{equation}
so that we recover a FRW-like metric in the standard form, meaning $g_{00}=-1$. It is clear
that the physical clock reduces to the momentum one for large $t$. In fact, its functional
form in terms of $t$ is very complicated, but its plot is easy to understand:
\begin{figure}[H]
\begin{centering}
\includegraphics[scale=0.3]{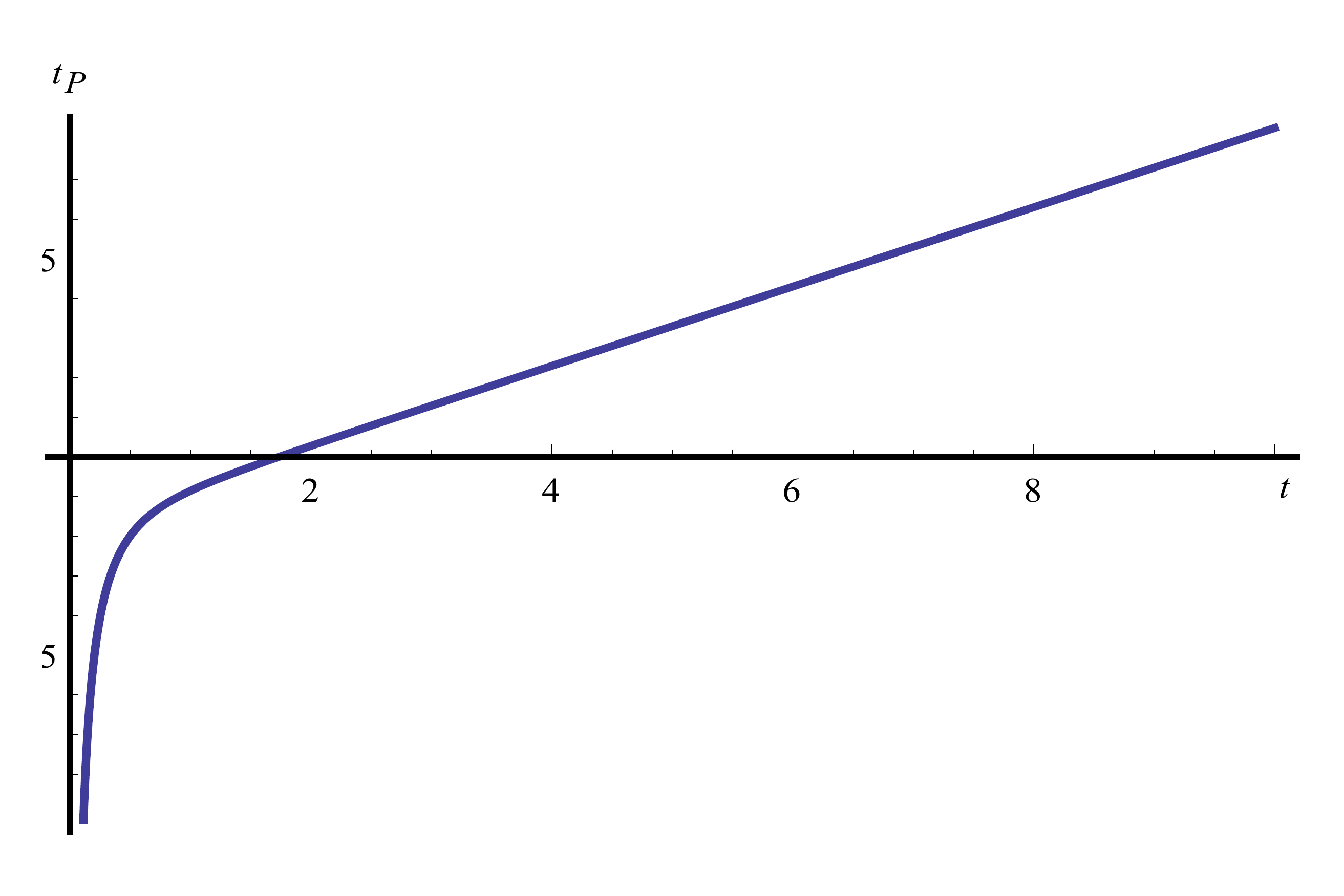}
\par\end{centering}
\caption{Physical clock as a function of the time coordinate. The physical clock goes from $-\infty$ to $\infty$ for $t\in(0,\infty)$.}
\end{figure}

The non-trivial solution for the scale factor (\ref{eq:non_trivial}) can also be plotted in terms of the physical clock (we consider $D=4$ for simplicity),
\begin{figure}[H]
    \centering
    \includegraphics[width=8cm]{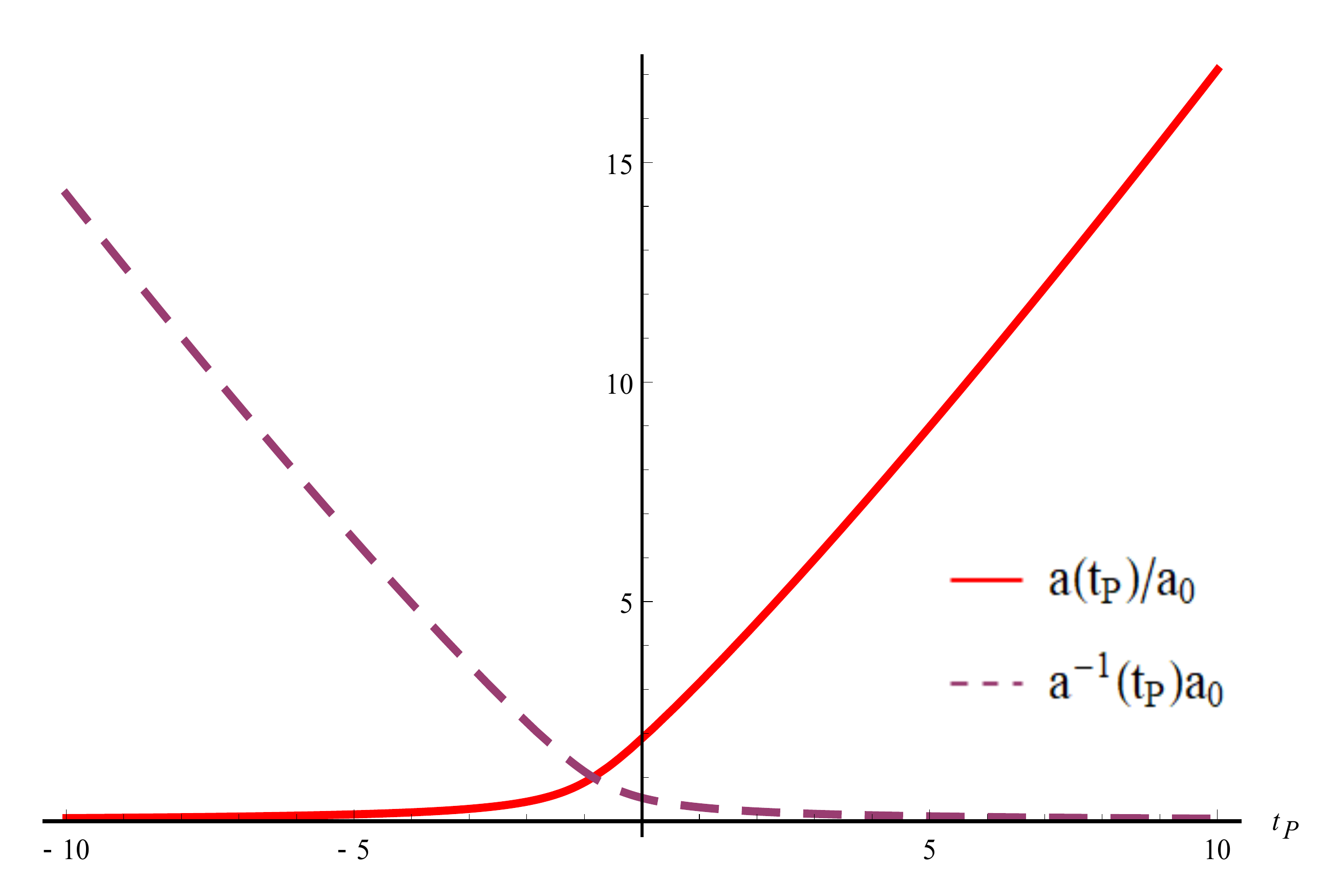}
    \caption{The scale factor goes to zero only at $t_p \rightarrow -\infty$. Similarly its inverse goes to zero when $t_p \rightarrow \infty$.}
    \label{fig2}
\end{figure}

Finally, the metric looks like
\begin{equation}
ds^{2} \, = \, -dt_{p}^{2}+a^{2}\left(t_{p}\right)\sum_{i=1}^{D-1}dx^i dx^i+a^{-2}\left(t_{p}\right)\sum_{j=1}^{D-1}d\tilde{x}^j d\tilde{x}^j,
\end{equation}
as expected. It is important to realize that this effective $(2D-1)$-dimensional geometry reduces
effectively to a $D$-dimensional one for large $t_p$, the momentum sector, and analogously
for large negative $t_p$, the winding sector. We also observe that there should be a $D$-dimensional
slice that has its volume bounded from below for all $t_p$. This slice is the physical
geometry where we live and which is accessible by physical rulers and clocks (which are always
given by the corresponding light modes).

\section{Conclusions and Discussion}

We have studied the geodesics corresponding to point particle motion
in Double Field Theory. We derived the general equations of motion,
and then considered the special case of a cosmological background
with vanishing antisymmetric tensor field. We
argued that in this context the geodesics of point particle motion are
complete, provided we measure the motion in terms of a physical time
which reflects the T-duality symmetry of the setup. Our result provides
further support for the expectation that cosmological singularities are
resolved in superstring cosmology. Then, we also considered a geometrization of
this prescription within the framework of DFT, resulting in a cosmological solution which
is singularity-free.

\section*{Acknowledgement}
\noindent

We thank Subodh Patil for suggesting to us to consider point particle
geodesics in a Double Field Theory background, and for many consultations
along the way. The research at McGill is supported in
part by funds from NSERC and from the Canada Research Chair program.
Two of us (RB and GF) wish to thank the Banff International Research
Station for hosting a very stimulating workshop ``String and M-theory geometries: Double Field Theory,
Exceptional Field Theory and their Applications" during which some of the
ideas presented here were developed. GF acknowledges financial support from
CNPq (Science Without Borders). RC thanks financial support by the SARChI NRF grantholder. AW and RC are supported by the South African Research Chairs Initiative of the Department of Science and Technology and the National Research Foundation of South Africa. Any opinion, finding and conclusion or recommendation expressed in this material is that of the authors and the NRF does not accept any liability in this regard. All the authors also thank financial support from  IRC - South Africa - Canada Research Chair Mobility Initiative, Grant No 109684.

\end{document}